# Active Distribution System Coordinated Control Method via Artificial Intelligence

Matthew Lau
Kayla Thames
Sakis Meliopoulos
School of Electrical and Computer Engineering
Georgia Institute of Technology
Atlanta, USA

*Abstract*—The increasing deployment of end use power resources in distribution systems created active distribution systems. Uncontrolled active distribution systems exhibit wide variations of voltage and loading throughout the day as some of these resources operate under max power tracking control of highly variable wind and solar irradiation while others exhibit random variations and/or dependency on weather conditions. It is necessary to control the system to provide power reliably and securely under normal voltages and frequency. Classical optimization approaches to control the system towards this goal suffer from the dimensionality of the problem and the need for a global optimization approach to coordinate a huge number of small resources. Artificial Intelligence (AI) methods offer an alternative that can provide a practical approach to this problem. We suggest that neural networks with self-attention mechanisms have the potential to aid in the optimization of the system. In this paper, we present this approach and provide promising preliminary results.

List of Abbreviations
A/C: Air conditioning
ADS: Active Distribution System
AI: Artificial Intelligence
BiLSTM: Bidirectional Long-Short Term
DER: Distributed Energy Resource
DG: Distributed Generation
HV: High Voltage
LSTM: Long-Short Term Memory
MPC: Model Predictive Controller
MSE: Mean Squared Error
MV: Medium Voltage
NARX: Nonlinear AutoRegressive network with eXogenous inputs
NLP: Natural Language Processing
NN: Neural Network
OLTC: On-Load Tap Changers
PV: Photovoltaic
RNN: Recurrent Neural Network
TRU: Transforming Recurrent Unit
VAE: Variational Autoencoder
VAr: Volt.Amperes reactive

## I. INTRODUCTION

The advent of the smart grid has generated technologies for real time control of distribution systems. Legacy approaches to optimally control power systems, including distribution systems, entail formulating optimal power flow approaches to determine the optimal settings of various resources of the system. The active distribution system has many resources with time dependent values, such as storage systems, and systems with mechanical thermal inertia such as thermostatically controlled loads (A/C, space heating, etc.). This makes it necessary to optimize the system over a time period, thus requiring multi-stage optimization. The plethora of distributed energy resources makes the relevant optimization problem large-scale and computationally intractable. Artificial Intelligence (AI)-based approaches offer an alternative that can provide solutions in real time.

As the penetration of distributed generation (DG) technologies such as photovoltaic (PV) cells, wind turbines, and other renewable energy sources grow, formerly passive distribution systems have transitioned to an active distribution system (ADS) with multidirectional power flow. To address the effects of DG units on power flow, various control systems have been adopted. [1] effectively deploy the reactive power of the DGs to keep ADS voltages within established parameters. [2] proposes primary and secondary control algorithms that adjust the voltage of a PV inverter. Similar to [1], the model used for the case study restricts the efficiency of the proposed control algorithms as it only models a five-home residential area connected to a grid. The distribution model used in the study limits the effectiveness of the applied control method as the DGs are not connected to the power grid in the form of a microgrid [3]. A voltage regulation problem was presented as an optimization problem in [4], with the goal of determining the ideal tap setting of a voltage regulator regulating several feeders. However, the interaction between the DG units and the on-load tap changers (OLTC) was not accounted for, which could result in incorrect tap setting or voltage violations.

AI methods have been proposed to tackle the complexity of such a problem. Specifically, neural networks (NN)s, which are typically more straightforward to implement, have achieved performance levels comparable to traditional optimization methods [5]. NNs are also able to obtain a balance between the accuracy and computational power of real-time model predictive controllers (MPC)s. In some cases, NNs have even outperformed traditional optimization methods; to better handle the complex constraints of an optimization problem, [6] demonstrated that using a NN as the controller managed to balance active power control and the volt/VAr control demand better than conventional control methods. The NN-MPC proposed in [7] achieved better load frequency control results for their simulated 2-area power system than fuzzy logic control, suggesting that a shift to using NNs would secure greater stability for power systems.

### A. Shortcomings of Present AI Approaches

Though powerful, the vanilla NNs built for the applications mentioned above do not consider the sequential nature of the data. While working with time-series data, data from previous time steps often affect data of future time steps. This key insight has often been used for state estimation and forecasting [16-18], but less so for MPC. If this time dependency can be modelled, there is potential for NN-MPCs to be even more accurate.

Outside the realm of large-scale power distributions, some work has been done to model the sequential nature of MPC data [8, 9]. Within the realm of power systems, Yang *et al.* proposed using recurrent neural networks (RNN)s with a Nonlinear AutoRegressive network with eXogenous inputs (NARX) structure for MPC to regulate building energy consumption [9]. Their NARX RNN MPC reduced the overall energy consumption, but the traditional MPC performed slightly better. Even within the realm of power systems, better AI methods are needed, especially for coordinated control of active distribution systems.

### B. Sequence Modelling in Machine Learning

Sequence modelling in machine learning is heavily driven by the field of Natural Language Processing (NLP). Previously, RNNs such as Long Short-Term Memory networks (LSTM)s were used to model the sequential nature of the text data. A pivotal turning point in NLP was the inspiration of the self-attention mechanism, which is now ubiquitous NLP [10, 13]. Self-attention contextualizes the text inputs of the NN, allowing the model to gain richer representations of every input word.

Building on the favorable results from using NNs as MPCs in active distribution systems, as well as drawing from inspiration from other fields like NLP, this paper contributes in the following ways:

1) We propose using NNs with self-attention for MPC. We compare other NN architectures and demonstrate its potential for MPC.

2) We use a high-fidelity simulator, which allows us to manipulate the settings of controls, to generate information about the various devices within an active distribution system.

## II. Problem Definition

Distribution grids have an average lifespan of 50 years; nevertheless, efficiency issues arise with technological advances and the introduction of new components to the power system. Some of the voltage and power quality issues that arise from the use of distributed energy resources (DER)s, namely: voltage fluctuations, reverse power flow (RPF), harmonics, local frequency oscillations, system stability, and protection issues [11]. This research aims to optimize voltage (which increases the overall efficiency of the ADS) as a means to monitor and control variable voltage. For this study, our primary focus is predicting the optimal regulator tap settings, PV phase angle, and capacitor settings. Machine learning approaches will be used to forecast the rate of change of states (voltage and current) for various states and control settings.

## III. Proposed AI Methods

This section presents the proposed method.

### A. MPC Formulation

The goal is to work towards a control system with desirable states that performs with a favorable voltage profile and efficiency. Intuitively, the input of sequence length two was modelled as such: given a previous state, its control settings and a desired state at the next time step, we seek to identify the optimal control settings that will shift the system to that desired state. For our proposed self-attention NN-MPC, we formulate the MPC to be:

$$u_i = f(u_{i-1}, x_{i-1}, x_i) \qquad (1)$$

where $f$ is the function of the NN used to model the MPC, $u_i$ refers to the vector of controls at time step $i$, and $x_i$ refers to the vector of system states at time step $i$.

### B. Overall AI Method

To test our proposed models, we compared the efficacy of deep self-attention models with other more conventional deep architectures, namely, the vanilla Dense model, LSTM and Bi-Directional LSTM (BiLSTM). Every architecture was tested across three categories: (A) architectures with a self-attention layer, (B) architectures with states and controls as input and (C) architectures with states as input. These categories will be detailed in the following sections. We also tested the 1-layer RNN model proposed by [9] with a 1-layer LSTM, and a multiheaded attention LSTM that has fewer layers than the ones in Category A.

Each layer from all NN models included a hyperbolic tangent activation (except for the last layer, which had a rectified linear unit activation) to include nonlinearities into each model. Under a mean-squared loss, the models were optimized using the Adam optimizer with an initial learning rate of 0.01 (coupled with learning rate annealing) with early stopping patience of 20 epochs monitoring the validation loss to allow for convergence while mitigating overfitting.

### C. Self-attention NN

Building on [13], we propose including a self-attention layer in our NN-MPC to better model the input data's sequential and potentially contextual nature. To demonstrate the effectiveness of the self-attention mechanism, we used not only the multiheaded scaled dot-product attention layers proposed by [13], but also show in Section IV the efficacy of the self-attention layer by using models with just a single-headed dot-product attention layer (without positional encoding, residual connections, and layer normalizations as per a traditional transformer architecture). We will refer to the set of models that have the single headed attention layer as Category A, and the subcategory of models with multiheaded attention as AM. This self-attention model is modelled after the Transforming Recurrent Units (TRU)s in [8], with a simpler architecture and training method, as well as an added self-attention layer at the input layer. We also switch around the recurrent units for LSTM, BiLSTM and Dense cells in our three models in Category A. Next, we detail our models' architecture.

A dot-product attention layer is built at the input layer of the sequence of states. This multiheaded attention layer is a function that takes in the sequence of states to output a rich representation of the sequence, which we will denote as $multihead$. Under the hood, the output is based on the attention scores $A_i$, which is the attention of $x_i$ (which is the state of the system at the $i^{th}$ time step of the sequence of $t$ states). We use Tensorflow to perform the following calculations [13]:

$$q_i = W_q x_i, \qquad k_i = W_k x_i, \qquad v_i = W_v x_i \qquad (2)$$

$$S_l = \frac{\exp\left(\frac{q_i \cdot k_l}{d_k}\right)}{\sum_j \exp\left(\frac{q_i \cdot k_j}{d_k}\right)} \qquad (3)$$

$$A_i = \sum_l S_l \times v_l \qquad (4)$$

$$head_j = [A_1, ..., A_t] \qquad (5)$$

$$multihead = concat(head_1, ..., head_H)W^O \qquad (6)$$

where $W_q$, $W_k$ and $W_v$ are the weight matrices for the query $q$, key $k$ and value $v$ that are learned in the training of the model. The query vector $q_i$ can be thought of as a question the $i^{th}$ state gets to ask a state to contextualize the representation of a state. The query is directed at a state's key vector $k_i$. The key-value pair[1] for each state may be interpreted as different representations of the state, where each key or value vector is a linear projection onto the span defined by its corresponding matrix. The dot products in Equation 3 measures the similarity between a question the model asks through the query vector and state $l$ through the key vector. The dot product is scaled down by a factor of $\sqrt{d_k}$, where $d_x \times d_k$ is the dimension of the key matrix for $x_i \in \mathbb{R}^{d_x}$, which helps to mitigate vanishing gradients if the softmax value is pushed to extreme points. The softmax score in Equation 3 weighs the importance of the value vector of the $l^{th}$ state, modelling the importance and relevance of the $l^{th}$ state in the calculation of the output attention score $A_i$. The attention score $A_i$ in Equation 4 is the final representation of the $i^{th}$ state, which considers the weighted contextual information from other states in the sequence. The combined attention scores are known as a "head", as shown in Equation 5. The multiheaded attention mechanism iterates this process over $H$ number of heads[2]. Having different heads allows for the model to learn different representations of the inputs per head, allowing for a richer representation of the inputs. The heads are then concatenated and multiplied with an output matrix $W^O$, which is learnt during training, as the input to the next layer of the NN, as seen in Equation 6. This final step combines the context and representations from each head. A more detailed explanation can be found in [13].

On top of the self-attention layer, the Category A models branch out into 3 slightly different models, each differing in the type of (custom) layer in the next layer (Figure 1): we tested a Dense layer (which models a simplified transformer architecture), an LSTM layer and a BiLSTM layer.

Separately, the controls from the previous time step are passed through a Dense layer, before being concatenated with the output of the LSTM/BiLSTM/Dense layer. The concatenated vector is then passed through 2 more hidden Dense layers before the final prediction of the controls which are needed to reach the last state in the given sequence of states. A visual representation is shown in Figure 1.

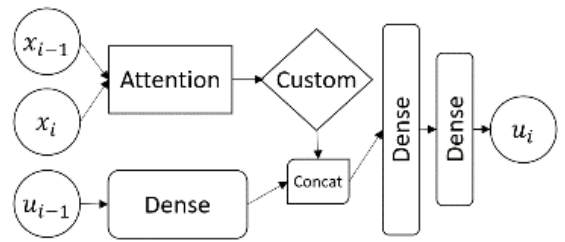

Figure 1: Proposed Category A NN Model Architecture

It should be noted, the simple multiheaded attention LSTM has 1 less Dense layer before the output and connects the controls $u_{i-1}$ directly to the concat layer without the Dense layer.

---

[1] A reference to databases.

[2] $H$ is a hyperparameter, and $H = 1$ for single-headed self-attention.

### D. Other Models Used

The second set of models, which we will denote as Category B, are models that are identical to those from Category A less the self-attention layer, as shown in Figure 2. Namely, we test the Dense, LSTM and BiLSTM models that have the sequence of states and the controls as the input. This comparison allows to better understand the effect of the self-attention layer.

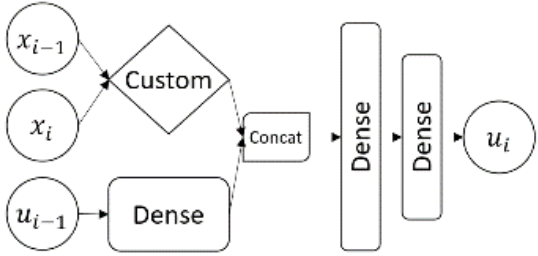

Figure 2: Category B NN Model Architecture

The third set of models, which we will denote as Category C, are models that have the architecture of Category B, but take in only the sequence of states as their input. The model architecture for Category C models can be visualized in Figure 2 without the $u_{i-1}$ control input. From Category C to B to A, we note the increasing complexity of the model. Hence, while more robust modelling of the interdependencies and sequential nature of the data is possible, there is also a risk of overfitting. We examine the effect of increasing the complexity of the models, especially in relation to our proposed addition of the self-attention layer.

### E. Training Method / Generation of Training Sets

We evaluate the model based on 2 metrics. Firstly, as presented in [14], we evaluate the various NN-MPCs with the mean-squared error (MSE) between the test set and the test prediction, as well as the accuracy of the predictions. Because there are controls which are continuous, we will be using MSE as our main metric. This evaluation ascertains the validity of the NN for predictive control. However, as seen in [19], the performance of NNs is dependent on its random initializations. To increase the reliability of our results, we experimented twice and averaged our results. We also trained our best performing (in terms of MSE) model obtained in the first round of training a second time (which was the Category AM LSTM).

Secondly, we assessed the efficiency of the system achieved by the predicted controls. This evaluation ascertains the validity of the NN as an MPC aimed at optimizing the efficiency of the system.

We also trained all NN models on the same computing power and considered the time taken to train and evaluate each model in the discussion section. Each model is trained with early stopping regularization. The time taken to train the model refers to the total time taken for the NN-MPC to converge. On the other hand, the evaluation time refers to the time taken for inference, which relates to the complexity of the model.

## IV. NUMERICAL EXPERIMENTS

Several illustrative examples of specific systems are presented. We use the high fidelity WinIGS software to design, simulate and perform analysis of the ADS.

### A. Example System Description

Figure 3 illustrates the model of the ADS developed for this study. For the predictive controls experiment, we primarily focused on a subsystem of the high voltage (HV) / medium voltage (MV) system that is part of the full ADS model (Figure 4). This system includes a substation, regulator, and a PV farm. The model was developed in the WinIGS-T program. In the simulation, the available controls included:

1) the controls of the PV system model that includes power tracking of the real power generated by controlling the voltage phase angle of the voltage source inverter and the reactive power control of the inverter,
2) the control settings of the capacitor (on or off), and
3) the control tap settings for the regulator (from 0.90 to 1.10).

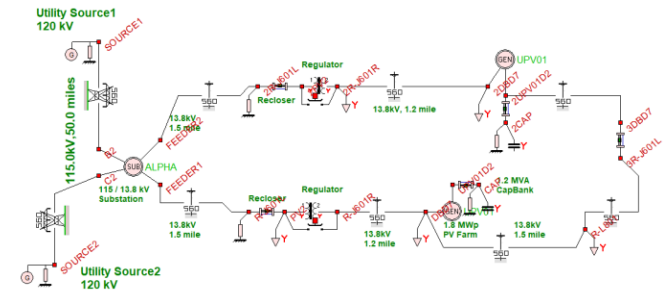

Figure 3: Complete Active Distribution Network

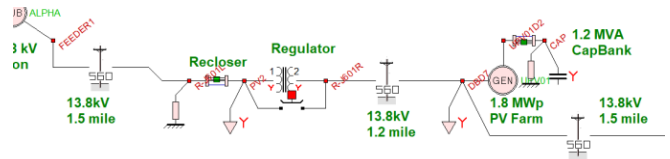

Figure 4: Basic Active Distribution Network

### B. Training Sets

Numerous combinations of settings were used to generate a total of 26 training sets. The training sets were then passed along to the Win-XFM software, which produced the harmonics for voltage waveforms. Generation of harmonic waves required the voltage waveforms of the multiple feeders in the system as inputs. The data collected from this software was then compiled and readied for use as the MPC model's input.

In total, 103,974 example cases were generated using the 26 training sets. The data was split into a 0.8-0.2 train-test split for the NN, and a further 20% of the training data was used as development data.

### C. Checking for Fidelity of the Simulation

To verify that the system is indeed performing accurately, we ran a power flow analysis on the distribution network using the WinIGS-F program, which has the capability to thoroughly describe the effects of a PV farm when integrated into a distribution system. As seen in Figure 5, a graphical report was generated from the analysis, which displays the phase voltage at each bus along the system. The phase voltage values for each node were within realistic expectations, which allowed for further analysis of the system. As discussed in the introduction, high penetration of DGs may result in overvoltage in ADSs caused by reverse power flow. Thus, reactive power optimization is employed to reduce network power losses and mitigate voltage violations [15]. The Power Balance Report, shown in Figure 6, provides the distribution system's real power and the reactive power totals as well as operational margins.

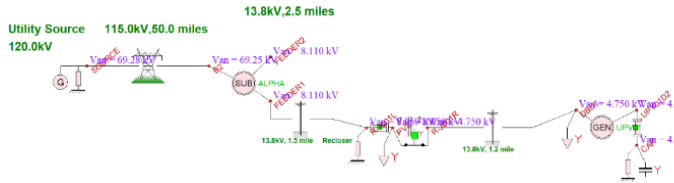

Figure 5: Phase voltages at each bus of the active distribution network

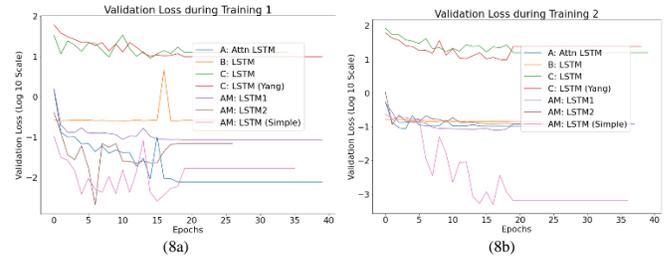

Figure 6: Report of Real and Reactive Power in system

We proceeded to generate voltage and current reports for all devices within the network. An example of a device's report of voltages and currents is shown in Figure 7. The reports of voltages and currents were vital as they informed us of the power losses experienced by all of the devices. The power losses for all devices within the network are summed and used to calculate the efficiency of the distribution network. To ensure that the ADS has high efficiency prior to the generation of data, we tested the network and calculated its efficiency. The calculated 96.1% efficiency is considered normal for an ADS. The data generated from the ADS by the WinIGS-T program is high fidelity as compared to the actual operation of the distribution system and therefore are considered realistic for the purposes of this study.

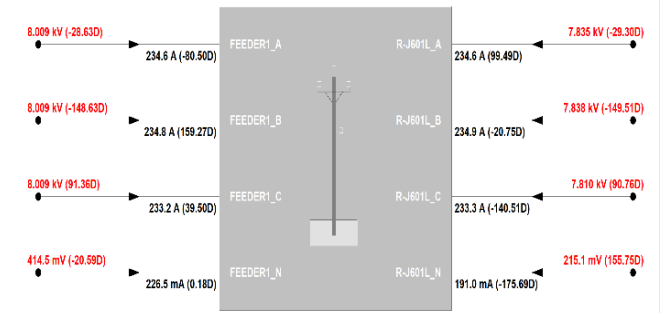

Figure 7: Report of Voltages and Currents for 1.5-Mile-Long Circuit

### D. Model Training

The training progression can be seen in Figure 8 where the logarithm of the validation loss after each epoch of training is plotted against the epoch count for the LSTM architecture.

Figure 8: Log validation loss against epochs (training round 1 & 2)

## V. RESULTS AND DISCUSSION

### A. Model Results and Discussion

As seen from Figure 9 and Table 1, the self-attention models (Category A/AM) had the best performance[3] by at least an order of magnitude for each architecture, followed by Category B models and then Category C models. When comparing Category A models with their Category B counterparts, an added self-attention layer (single or multiheaded) reduced test loss by 72.2% in the BiLSTM model, 75.9% in LSTM model and 99.5% in the Dense model (Figure 9). Special mention goes to the simple LSTM-layered multiheaded attention model inspired by [9], which outperformed the original 1-layer LSTM model proposed. On average, it tops the leaderboard, achieving the best overall test loss of 0.000474, which is a 99.8% decrease in test loss compared to its Category B LSTM counterpart. This result is significant because it suggests that the inductive bias of sequence modelling (through the self-attention mechanism) is far more vital than having a deeper NN architecture in achieving accurate predictions. In all cases, we have shown that the self-attention mechanism significantly improves the predictive capability of the NN. Furthermore, as seen in Table 1, this increase in performance is not at the cost of time used for training or inference.

---

[3] Measured by lowest test loss.

Table 1: Averaged Model Test Loss & Accuracy, Training and Evaluation Time, Sorted by Test Loss

| Model | Loss | Acc | Train Time | Eval Time |
|---|---|---|---|---|
| AM: LSTM (Simple) | 0.0016 | 98.0% | 352.6 s | 1.61 s |
| A: Attn Dense | 0.0169 | 97.9% | 189.0 s | 1.19 s |
| AM: Dense | 0.0446 | 98.3% | 200.0 s | 1.20 s |
| A: Attn LSTM | 0.0473 | 96.1% | 266.2 s | 1.54 s |
| AM: LSTM2 | 0.0609 | 96.1% | 245.7 s | 1.66 s |
| A: Attn BiLSTM | 0.0616 | 96.1% | 388.3 s | 1.78 s |
| AM: LSTM1 | 0.0813 | 94.3% | 387.5 s | 1.62 s |
| AM: BiLSTM | 0.0870 | 95.0% | 335.8 s | 1.78 s |
| B: LSTM | 0.1962 | 94.6% | 268.8 s | 1.44 s |
| B: BiLSTM | 0.2923 | 90.4% | 386.1 s | 1.71 s |
| B: Dense | 3.1901 | 86.2% | 192.3 s | 1.12 s |
| C: LSTM (Yang [9]) | 9.9850 | 94.7% | 285.9 s | 1.32 s |
| C: BiLSTM | 12.9730 | 93.0% | 797.9 s | 2.80 s |
| C: LSTM | 13.1191 | 91.6% | 472.4 s | 1.86 s |
| C: Dense | 42.7668 | 88.0% | 209.7 s | 1.11 s |

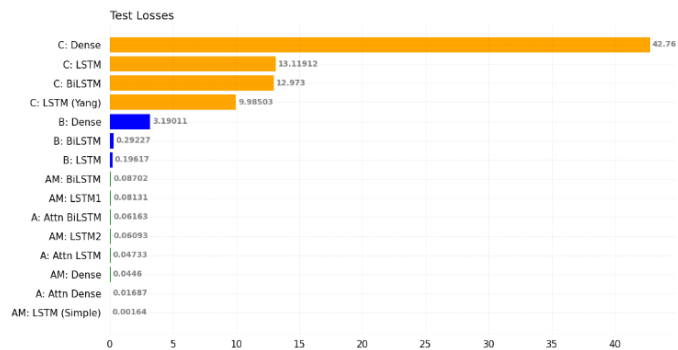

Figure 9: Averaged Test Losses for NN-MPCs

Our inclusion of the self-attention mechanism improves the model's ability to contextualize the inputs. This is in the spirit of passing in the control settings from the previous time steps to give the NN more context for its prediction of the next control setting. The advantage of our models over attentionless models is the ability to model the context of the situation for the model to provide a more suitable prediction; for instance, given some sequence of states, the self-attention mechanism helps to model which states bear more significance on the final prediction of controls. A higher weight for a state is an indication by the NN-MPC that the state is highly correlated to the control setting in the prediction. This increased interpretability of self-attention networks in MPC is explored in [12].

As perceptively inferred in [9], we note that simpler models do not underperform, but certainly have the potential to be more powerful – the LSTM NN modelled after Yang *et al.* [9] and the simple multiheaded attention LSTM bested the other models in their respective categories. A deeper network could diminish the gradients during backpropagation (termed as the vanishing gradient problem) which could have impacted convergence. Its over-parameterization could also have been overfitting the training data more than shallower networks do.

As we validated, the simplified transformer architecture (AM: Dense model) as outlined in [13] performs well. In addition, we have also shown that the self-attention mechanism works well with an LSTM layer, which adds another layer of modelling the sequential nature of the state data.

### B. Efficiency Analysis Results and Discussion

Since the LSTM architecture performed the best in their respective categories, we selected the Category AM LSTM (simple), the Category B LSTM and the Category C LSTM models to test their performance on driving up the efficiency of the power system, which was simulated and analyzed (alongside the control model without optimization) in WinIGS-F. The models acted as MPCs on three base control models, labelled as Control Model 0, 1 and 2. The base control models were initialized with the same phase angle but with differing regulator tap and capacitor settings. Figure 10 illustrates the efficiencies of base control models (without optimization) and the efficiencies of their predictive control models. To ensure the validity of the model, whenever the predicted control values were out of the parametrized range, they were rounded to the nearest feasible value.

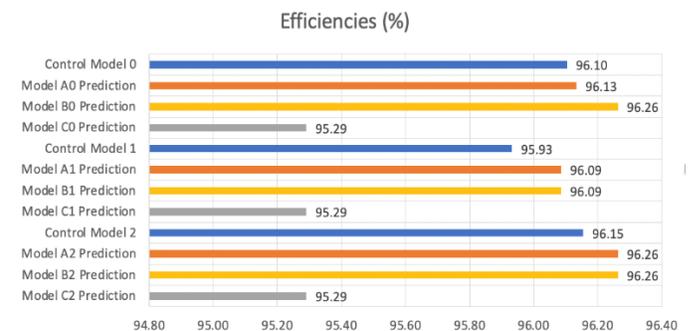

Figure 10: The Efficiencies of Base Control Models and Each of Their Respective Predicted Control Models

We note that for all 3 base control models, models A and B predicted controls which consistently optimized the power system, and model B's predictions resulted in a slightly greater efficiency than model A's predictions (for the first case). The increased efficiency should be surprising – when formulating the MPC problem, we sought to solve a sub-problem instead, which was to predict the optimal controls to get from one state to the next. Solving this sub-problem does not necessarily translate into solving the general problem of efficiency of the system over time; the model is trained to get from one state to another efficiently, but the model is not explicitly trained to know which state is the best to achieve the optimal operation of the power system. However, the fact that the Category A and B models achieved efficiencies better than base efficiency when trained on an auxiliary task hints that, with a more rigorous formulation of the problem, the self-attention model has much potential to be an effective MPC, given its superior accuracy in the theoretical predictions. Additionally, we validate the robustness of conventional sequence modelling methods, as we see the Category B LSTM achieving good efficiencies as well.

### C. Extensions

Our path forward would focus on formulating a more robust MPC that more directly tackles the problem of power systems optimization. Instead of solely taking in information about the states and their corresponding control inputs, the system can be more effectively optimized by having the NN-MPC also consider the efficiency of the system while the system runs. For instance, rather than solely an MSE loss, modelling the efficiency of the system in the objective function of the NN-MPC has the potential to increase the effectiveness of the NN-MPC as not just a NN, but also an MPC aimed towards optimization. Reinforcement learning could work in tandem with our model as well, and perhaps be able to model even longer-term efficiencies with an MPC controller that interacts directly with the simulation. A possible extension would be to add more states to the input to strengthen the Markov assumption in reinforcement learning that the "present state" (which are the previous few time steps in this case) encapsulates all the information needed to predict future states.

More work can also be done in analyzing the effect of model architectures. For one, it could be interesting to understand how certain inductive biases interplay with the pros of adding depth to a NN, and possibly even with using the full transformer architecture instead of just the self-attention layer. Probabilistic models such as variational autoencoders (VAE)s can also be included in pre-training, which could potentially increase the robustness of noisy physical systems like distribution systems.

## VI. CONCLUSIONS

The proposed deep learning architecture with self-attention outperforms the other conventional deep architectures in MPC by predicting more accurate controls. Additionally, our self-attention NN-MPC was able to optimize the system without an explicit formulation of the MPC problem aimed at voltage optimization. We foresee that, with a more direct MPC formulation to account for voltage optimization, the self-attention NN-MPC has the potential to provide reliable coordinated control for an active distribution system with large numbers of distributed energy resources.

## VII. ACKNOWLEDGMENT

The authors acknowledge the DoE/SETO support via award DE-EE0009020. Models used to perform research for this paper and more detailed data visualizations can be found in the repository at https://github.com/BPod123/WinIGS-Tool-it/tree/master/Optimization.

## IX. BIOGRAPHIES

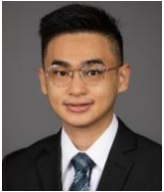

**Matthew Lau** is pursuing a B.S. degree in Mathematics at Georgia Institute of Technology, doing research with the PSCAL Research Group at Georgia Institute of Technology. His research interests include applications of machine learning in noisy data, such as in cyber-physical systems and anomaly detection.

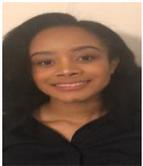

**Kayla Thames** received her B.S. degree in Civil Engineering from Morgan State University, Maryland, USA, in 2021. She is currently pursuing the Ph.D. degree in Electrical and Computer Engineering as part of the PSCAL Research Group at Georgia Institute of Technology. Her research interests include power system protection, renewable energy integration, and machine learning.

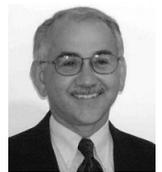

**A. P. Sakis Meliopoulos** (M '76, SM '83, F '93) was born in Katerini, Greece, in 1949. He received the M.E. and E.E. diploma from the National Technical University of Athens, Greece, in 1972; the M.S.E.E. and Ph.D. degrees from the Georgia Institute of Technology in 1974 and 1976, respectively. In 1971, he worked for Western Electric in Atlanta, Georgia. In 1976, he joined the Faculty of Electrical Engineering, Georgia Institute of Technology, where he is presently a Georgia Power Distinguished Professor. He is active in teaching and research in the general areas of modeling, analysis, protection, and control of power systems. He has made significant contributions to power system grounding, harmonics, protection and reliability assessment of power systems. He is the author of the books, Power Systems Grounding and Transients, Marcel Dekker, June 1988, Lightning and Overvoltage Protection, Section 27, Standard Handbook for Electrical Engineers, McGraw Hill, 1993. He holds three patents and has published over 300 technical papers. In 2005 he received the IEEE Richard Kaufman Award and in 2010 he received the George Montefiore Award from the Montefiore Institute, Belgium. Dr. Meliopoulos is the Chairman of the Georgia Tech Protective Relaying Conference, a Fellow of the IEEE and a member of Sigma Xi.